\documentclass[fleqn,10pt]{wlscirep}
\usepackage{bm}
\usepackage{amsmath}
\usepackage{graphicx}
\usepackage{epstopdf}
\usepackage[version=3]{mhchem}
\usepackage{pdfpages}
\title{Novel Electron-Phonon Relaxation Pathway in Graphite Revealed by Time-Resolved Raman Scattering and Angle-Resolved Photoemission Spectroscopy}

\author[1]{Jhih-An Yang}
\author[1]{Stephen Parham}
\author[1,2]{Daniel Dessau}
\author[1,*]{Dmitry Reznik}
\affil[1]{Department of Physics, University of Colorado at Boulder, Boulder, CO 80309}
\affil[2]{Center For Experiments on Quantum Materials, University of Colorado, Boulder, CO 80309}

\affil[*]{Dmitry.Reznik@colorado.edu}

\begin{abstract}
Time dynamics of photoexcited electron-hole pairs is important for a number of technologies, in particular solar cells. We combined ultrafast pump-probe Raman scattering and photoemission to directly follow electron-hole excitations as well as the G-phonon in graphite after an excitation by an intense laser pulse. This phonon is known to couple  relatively strongly to electrons. Cross-correlating effective electronic and phonon temperatures places new constraints on model-based fits. The accepted two-temperature model predicts that G-phonon population should start to increase as soon as excited electron-hole pairs are created and that the rate of increase should not depend strongly on the pump fluence. Instead we found that the increase of the G-phonon population occurs with a delay of $\sim$65 fs. This time-delay is also evidenced by the absence of the so-called self-pumping for G phonons. It decreases with increased pump fluence. We show that these observations imply a new relaxation pathway: Instead of hot carriers transferring energy to G-phonons directly, the energy is first transferred to optical phonons near the zone boundary K-points, which then decay into G-phonons via phonon-phonon scattering. Our work demonstrates that phonon-phonon interactions must be included in any calculations of hot carrier relaxation in optical absorbers even when only short timescales are considered.
\end{abstract}
\begin{document}

\flushbottom
\maketitle

\thispagestyle{empty}

\section*{Introduction}

The ultrafast dynamics and relaxation of hot carriers and phonons in a variety of materials investigated by pump-probe techniques has attracted a lot of attention in the last decade \cite{nmat3757,PhysRevLett.111.027403,srep00064,PhysRevLett.115.086803,0953-8984-27-16-164206,ncomms2987,PhysRevLett.102.086809,PhysRevB.83.153410,nphys2564,PhysRevLett.108.167401,apl1.2809413,nl1024485,apl/96/8/10.1063/1.3291615,PhysRevLett.95.187403,PhysRevB.85.125413}. Rapid development of pulsed lasers combined with increased utilization of optical absorbers \cite{nphoton.2010.186,nphoton.2010.40,Gabor648} stimulated spectroscopic investigations of materials on picosecond and sub-picosecond timescales under optically driven non-equilibrium conditions. Time-resolved experiments use intense laser pulses to create non-equilibrium states in materials and track their relaxation in real time. They provide information about unoccupied electronic states, electron-electron (e-e) interactions \cite{ncomms2987,PhysRevLett.115.086803,nmat3757}, electron-phonon (e-ph) coupling \cite{PhysRevB.77.121402,PhysRevLett.95.187403,nphys2564}, phonon-phonon (ph-ph) coupling \cite{PhysRevB.80.121403,PhysRevB.83.205411}, etc. Here we show that combining time-resolved photoemission with time-resolved Raman (TRR) \cite{PhysRevLett.54.2151} scattering provides insights that are much more difficult to obtain otherwise. Angle-resolved photoemission spectroscopy (ARPES) views the process from electron perspective tracking electron-hole (e-h) excited state occupation as a function of time. Raman scattering used as a probe reveals the phonon perspective providing vibration frequency, lifetime, and occupation number of Raman-active phonons. The fundamental physical law of detailed balance allows extracting the phonon occupation number $n$ from each Raman-active phonon directly from the ratio of the energy loss (Stokes) and energy gain (anti-Stokes) one-phonon Raman scattering intensities \cite{hayes2012scattering}. The anti-Stokes side, where the phonon intensities are proportional to $n$, is especially sensitive. 

Up to now the data have been described in terms of the two-temperature (2T) model where hot electrons thermalize with a few "hot" phonons relatively quickly (few hundred fs) via strong e-ph coupling \cite{PhysRevLett.99.197001,apl/96/8/10.1063/1.3291615,PhysRevLett.111.027403}. In this description, hot electrons and hot phonons are characterized by their temperatures. Other e-ph and ph-ph interactions are assumed to be much weaker resulting in the whole system reaching the final thermal equilibrium at much longer picosecond timescales. 

Our investigation focused on graphite as a model electron-phonon system. TRR as well as trARPES has been previously applied to graphene, graphite, and carbon nanotubes \cite{PhysRevLett.100.225503,PhysRevB.80.121403,nl802447a,PhysRevB.81.165405,nl301997r,PhysRevB.83.205411,nmat3757,srep00064,PhysRevLett.111.027403}, but they have never been applied to one compound simultaneously as we have done in this paper. G-phonons in graphite and related materials are believed to play the role of "hot" phonons in the 2T description due to strong coupling to electrons. We observed G-phonon population (data points in Fig. \ref{Ph_T}\color{blue}(a) \color{black}) as a function of time in the sub-picosecond region by Raman scattering and electron-hole population (black points in Fig. \ref{Ph_T}\color{blue}(b) \color{black}) by ARPES with nearly the same time-resolution and pump fluence. The unexpected result is that G-phonon generation is delayed by about 65 fs from the prediction of the 2T model (green dashed line in Fig. \ref{Ph_T}\color{blue}(a)\color{black}). Furthermore, the ARPES measurements  show that the e-h pairs decay faster than expected from direct coupling to the G-phonon, and the G-phonon generation time constant is reduced with a higher pump power in contradiction with the 2T model. This suggests that hot carriers couple to G-phonons not directly but via an intermediate excitation that is not Raman active (red lines in Fig. \ref{Ph_T}\color{blue}(a,b) \color{black}). We argue that K-phonons play this role (Fig. \ref{Ph_T}\color{blue}(c)\color{black}). Our results agree with LDA calculation that predicts that e-h pairs predominantly decay into K-phonons \cite{PhysRevB.78.081406}.

\section*{Experimental Results}
\begin{figure}[t]
\includegraphics[width=0.9\textwidth]{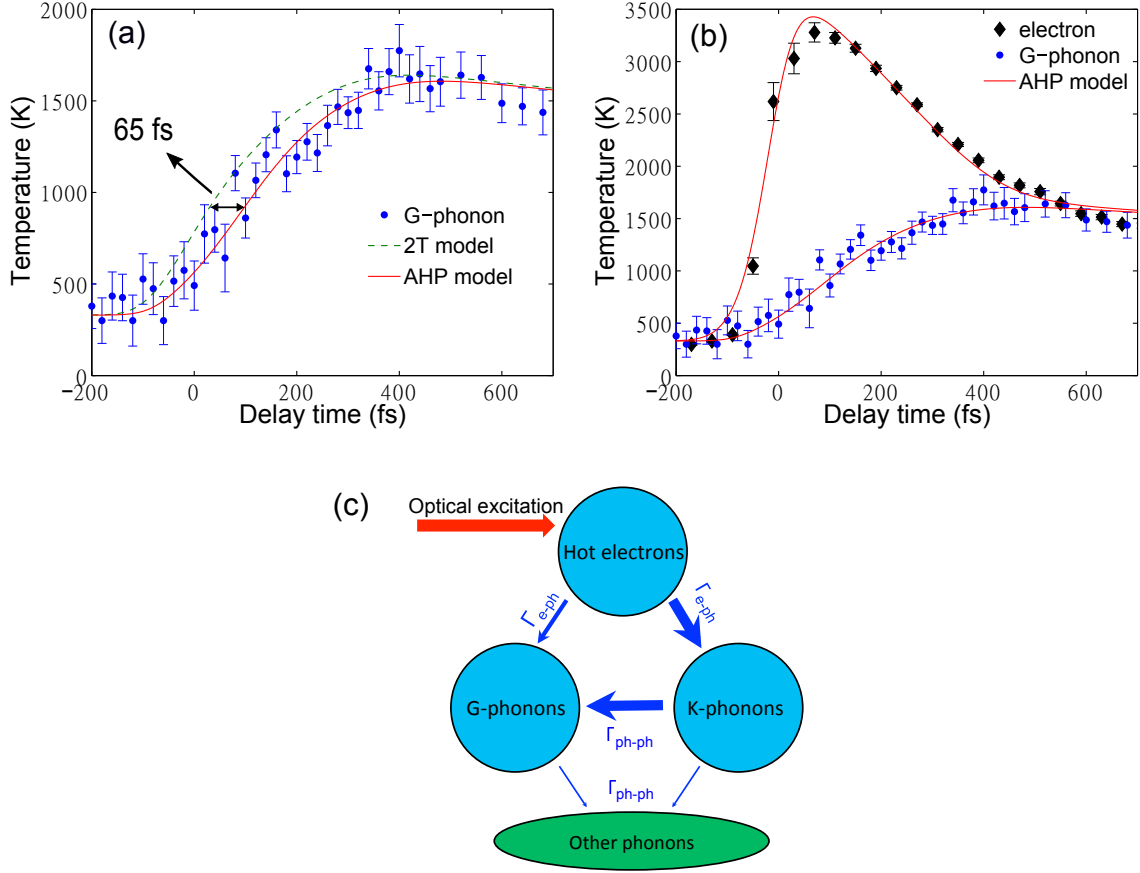} 
\centering
\caption{(a) G-phonon temperature (blue dots) inferred from the Raman Stokes/anti-Stokes ratio and the fits from the 2T model (green dashed line) and the anharmonic hot phonon (AHP) model (red solid line). The time shift between the green and red curves is about 65 fs. (b) Electronic temperature measured by ARPES (black) and G-phonon temperature measured by Raman scattering (blue, same data points as in (a)). The red curves are the fits from the AHP model. The pump fluences for Raman and ARPES are 0.15 $mJ/cm^2$ and 0.13 $mJ/cm^2$ respectively. (c) Sketch of energy transfer during the hot carrier relaxation. Thicker arrows represent faster processes.\label{Ph_T}}
\end{figure}
\begin{figure}[t]
\centering
\includegraphics[width=1\textwidth]{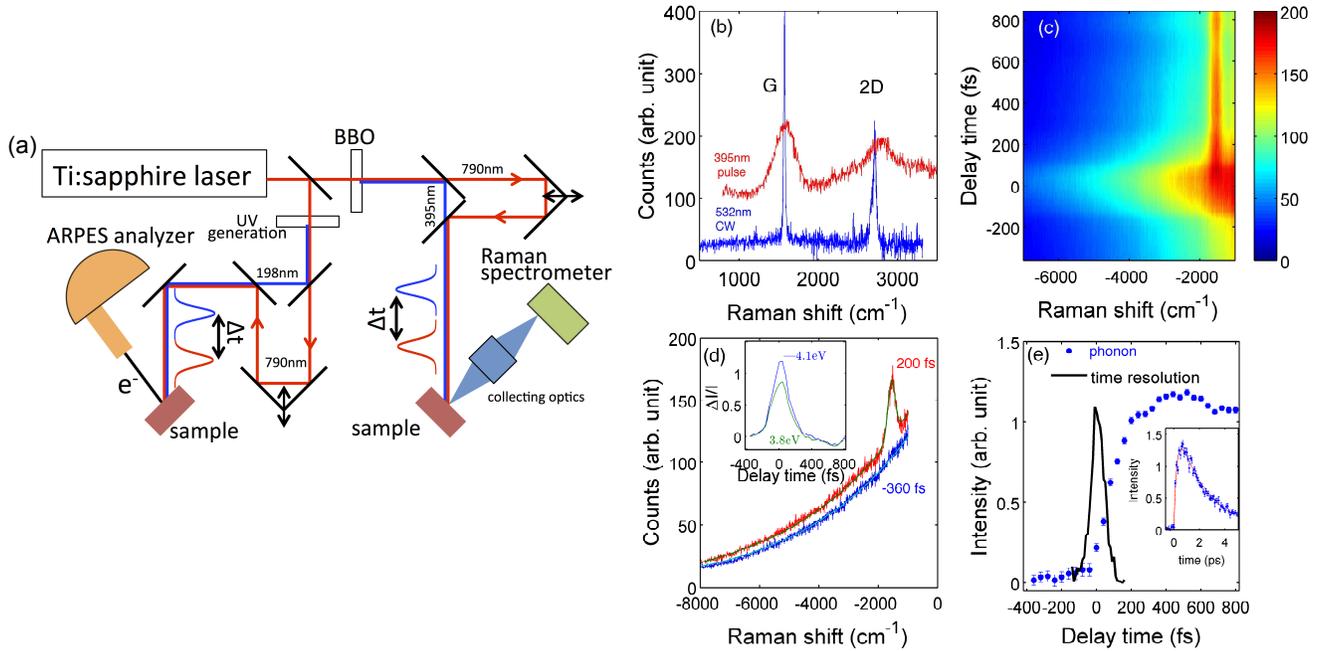} 
\caption{(a) A schematic of the combined TRR and trARPES system. The laser is converted into 395 nm or 198 nm using one or three non-linear crystal $\beta$-BaB$_2$O$_4$ (BBO). (b) Stokes Raman spectra measured with 532 nm CW laser (blue) and 395 nm pulse laser (red). The phonon peaks measured by the 395 nm pulse sit on a luminescnence background due to hot carriers. (c) Anti-Stokes Raman spectra as a function of delay time. G phonons (-1580 cm$^{-1}$) appear after an optical excitation and decay gradually. The pump fluence is 5 $mJ/cm^2$. (d) Anti-Stokes Raman spectra at $t$ = -360 fs (blue) and $t$ = 200 fs (red). The solid lines represent fits with Plank's law plus a Gaussian function. The luminescence background can be described by Plank's law in terms of absolute energy instead of relative Raman shifts \cite{PhysRevLett.105.127404}. The inset shows the background change at various delay times at 4.1 eV and 3.8 eV (selected absolute energy) normalized to the intensity at negative time.  (e) Temporal evolution of phonons on the anti-Stokes side (blue dots). The thick black line is the resolution curve. The inset shows the evolution of phonons from $t$ = -500 fs to $t$ = 5 ps measured with larger time-steps. Line represents the fit to a biexponential function convolved with the instrument response. \label{long_time}}
\end{figure}
\begin{figure}[t!]
\centering
\includegraphics[width=0.5\textwidth]{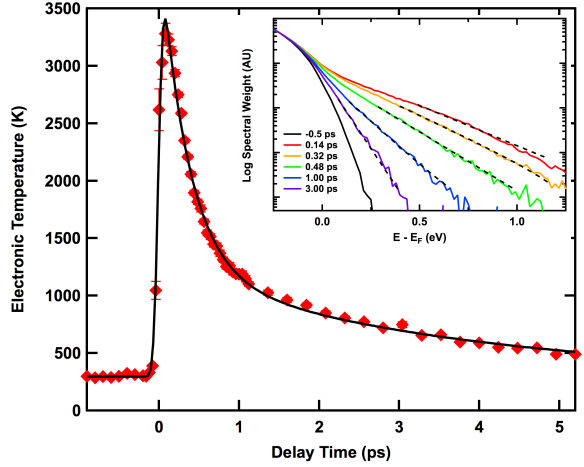} 
\caption{Electronic temperature extracted from trARPES data and the fit of a bi-exponential decay. The inset shows momentum-integrated ARPES spectra (spectral weight on log scale) at various delay times. The dashed black lines are fits for the selected high energy tails, which were used to extract the electronic temperature of the main panel. \label{trARPES}}
\end{figure}

\begin{figure}[t!]
\centering
\includegraphics[width=1\textwidth]{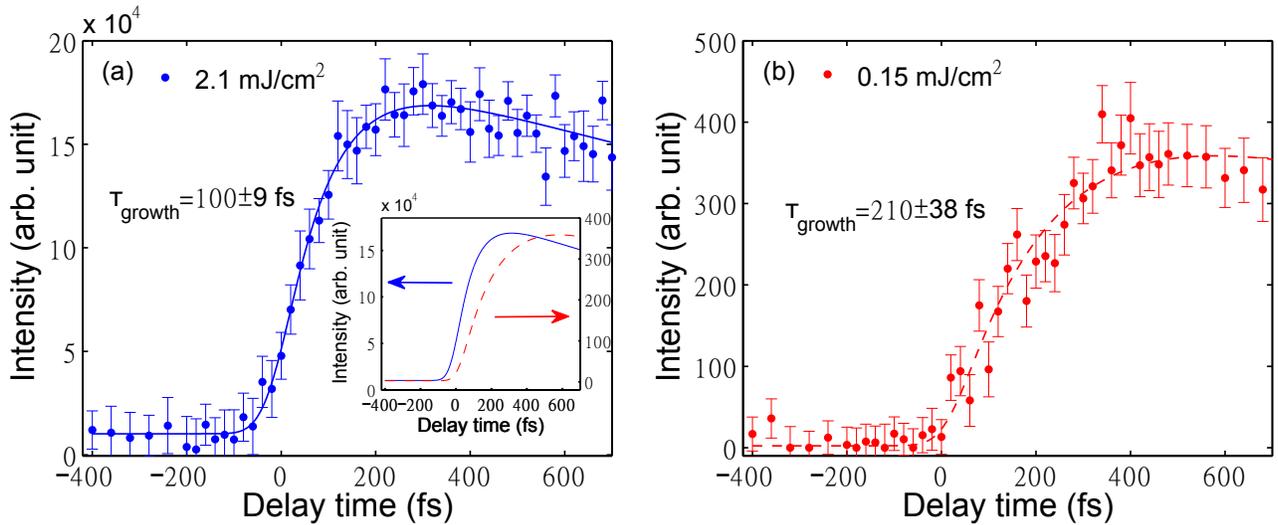} 
\caption{G phonon intensity on the anti-Stokes side at various delay times with pump fluence of (a) 2.1 $mJ/cm^2$ (blue) and (b) 0.15 $mJ/cm^2$ (red, same data points as in Fig. \ref{Ph_T}\color{blue}(a) \color{black}). The solid and dashed curves are the fits using a biexponential function $-exp[-(t-t_0)/\tau_{growth}]+exp[-(t-t_0)/2400]$. The fitting parameter $\tau_{growth}$ is $100\pm 9$ fs for (a) and $210\pm 38$ fs for (b), and $t_0$ is $-23\pm7$ fs for (a) and $10\pm 13$ fs for (b). The inset shows overplotting of data with different pump powers. \label{fluence_dep}}
\end{figure}

\begin{figure}[t!]
\centering
\includegraphics[width=0.5\textwidth]{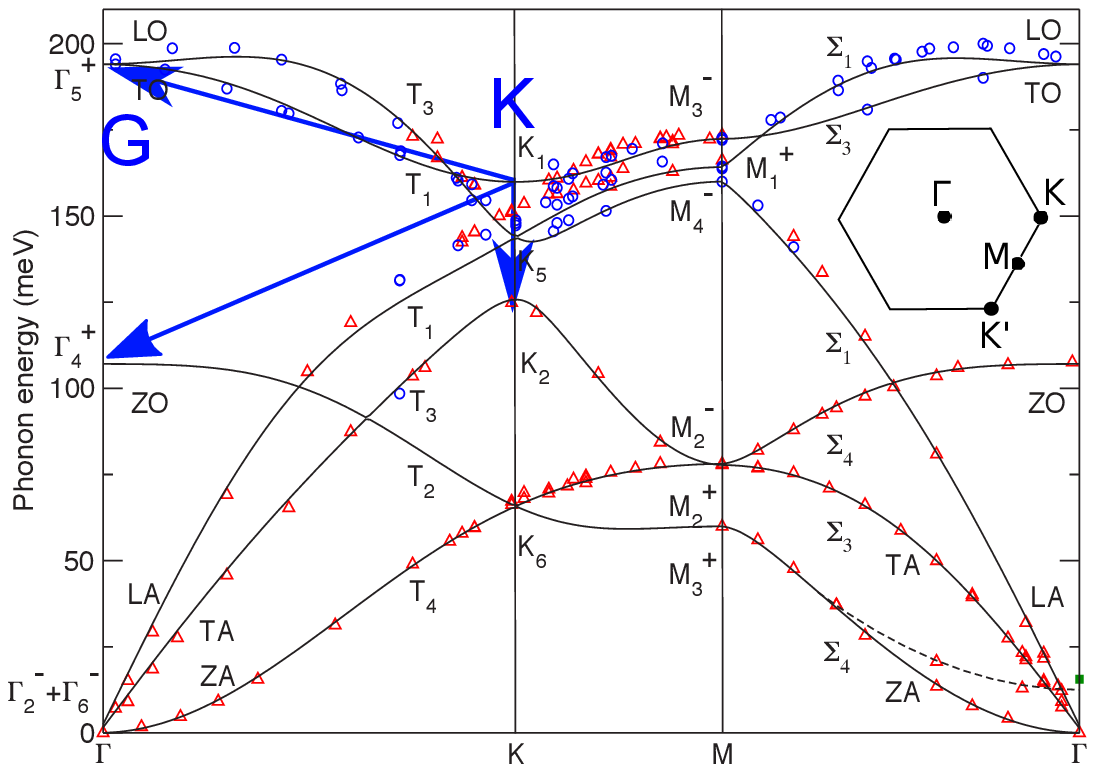} 
\caption{Phonon dispersions and possible anharmonic decays for K-phonons by 4-phonon scattering (blue arrows). The inset shows the schematic of the first Brillouin zone. Due to energy and momentum conservation, two K phonons can decay into two phonons at $\Gamma$ points or decay into one phonon at $\Gamma$ points plus one phonon at K points. Figure adapted with permission from Ref. \cite{PhysRevB.76.035439} copyrighted by the American Physical Society.  \label{dispersion} }
\end{figure}

The Raman spectrum of graphite measured with TRR setup shown in Fig. \ref{long_time}\color{blue}(a)\color{black} has two main features: the G peak at $\sim$1580 cm$^{-1}$ and the 2D peak at $\sim$2700 cm$^{-1}$ (Fig. \ref{long_time}\color{blue}(b)\color{black}). The G peak is the zone center $E_{2g}$ phonon. The 2D peak is due to the two-phonon scattering at K points called double resonance \cite{PhysRevLett.97.187401,nnano.2013.46,Ferrari200747}. These peaks measured by the 395 nm ultrafast probe pulse sit on a luminescence background and have large linewidths due to the finite bandwidth of the pulse. The absence of the defect-induced D peak ($\sim$1350 cm$^{-1}$) indicates good quality of our sample. Fig. \ref{long_time}\color{blue}(c) \color{black} displays the anti-Stokes spectra at various delay times. We fit the phonon spectrum at each time delay with a Gaussian (Fig. \ref{long_time}\color{blue}(d)\color{black}) and then extracted the integrated intensity as a function of time (Fig. \ref{long_time}\color{blue}(e) \color{black} blue dots and inset). 

The time evolution of the G-phonon can be fit with two exponential functions $-exp[-(t-t_0)/\tau_{growth}]+exp[-(t-t_0)/\tau_{decay}]$ convolved with the instrument response. The time constant $\tau_{growth}$ is 220 $\pm$ 20 fs for the quick buildup of G-phonons and $\tau_{decay}$ is 2400 $\pm$ 83 fs for the slow decay (Fig. \ref{long_time}\color{blue}(e) \color{black} inset). The fitting parameter $t_0=2 \pm 15$ fs. The G-phonon temperature is determined from the G peak intensity ratio on the Stokes and anti-Stokes sides $I_S/I_{AS}=exp(\hbar\Omega/k_{B}T_G)$, where $\hbar\Omega$ is the G-phonon energy, $k_B$ is the Boltzmann's constant, and $T_G$ is the phonon temperature (Fig. \ref{Ph_T}\color{blue}(a) \color{black}). $T_G$ is over 3000 K (over 1650 K for pump fluence 0.15 $mJ/cm^2$ as shown in Fig. \ref{Ph_T}\color{blue}(a) \color{black}) around $t$ = 400 fs. It decreases to room temperature with a time constant of 2.4 ps, which agrees with previous results \cite{PhysRevB.80.121403,PhysRevB.83.205411}. The slow decay has been attributed to the energy transfer from G-phonons to other phonon modes via anharmonic coupling \cite{PhysRevLett.100.225503,PhysRevLett.99.176802}. The apparent oscillation around $t=600$ fs in Fig. 2d might be due to imperfect laser power stability. We did not observe this oscillation in other datasets.

Fig. \ref{trARPES} shows electronic dynamics probed by trARPES. Electronic temperature $T_{el}$ was determined by fitting the slope of the high energy tail in the momentum-integrated ARPES spectra (Fig. \ref{trARPES} inset). $T_{el}$ increases rapidly after an optical excitation and reaches 3400 K at $t$ = 100 fs. $T_{el}$ fits well to a bi-exponential decay with decay times of $\tau_{fast}=0.35$ ps and $\tau_{slow}=3.5$ ps. The decay of $T_{el}$ has a similar timescale to the one observed by previous photoemission experiments\cite{nmat3757,PhysRevLett.111.027403,0953-8984-27-16-164206}. We evaluated the ratio of the electronic heat capacity to G-phonon heat capacity (see supplementary) from the electronic temperature and G-phonon temperature. It is one of the parameters in our theoretical model as will be discussed later. The 2D Raman peak also reflects electronic temperature, though it is less precise than trARPES results due to self-pumping and greater statistical error (see supplementary).

Fig. \ref{fluence_dep} shows G-phonon dynamics with pump fluence 2.1 $mJ/cm^2$ and 0.15 $mJ/cm^2$ respectively. The phonon dynamics are similar for these two pump powers. Maximum phonon temperature decreases with lower pump fluence as expected. The generation of G phonons is slower with low pump fluence ($\tau_{growth}=210$ fs) than for the high pump fluence ($\tau_{growth}=100$ fs).

We now briefly discuss the electronic background in the Raman spectra, which has been assigned to hot carrier luminescence  \cite{PhysRevLett.105.127404}. Fig. \ref{long_time}\color{blue}(d) \color{black} inset shows that the background starts to increase at $t$ = -200 fs and reaches its maximum at $t$ $\sim$ 0 fs, when the G-phonons begin to appear. The background then decreases to the initial level at $t$ = 300 fs, while the phonon intensity keeps increasing from $t$ = 0 fs to $t$ = 300 fs. This behavior reflects changes in electronic temperature and is consistent with previous reports \cite{PhysRevLett.105.127404,PhysRevB.82.121408,PhysRevB.82.081408,nl504176z}. However, it is not possible to read out instantaneous electronic temperatures from the observed intensities.

\section*{Discussion}
Our ultrafast pump pulse generates hot electrons and holes, which quickly thermalize via e-e interaction on femtosecond timescales \cite{PhysRevLett.111.027403,nmat3757}. These e-h pairs decay into phonons with strong e-ph coupling on longer timescales \cite{apl1.2809413,apl/96/8/10.1063/1.3291615,PhysRevLett.95.187403,PhysRevB.85.125413}; The hot phonons in turn slowly decay into other phonons, which carry energy out of the system \cite{PhysRevB.80.121403}.  Here we focused on graphite where there are no experimental complications due to the substrate and sample size issues. Our results also apply to graphene since its high energy electronic structure is similar. The low energy states with energy less than the hot-phonon frequency are less relevant for the phonon dynamics but are relevant for other physics such as high electric and thermal conductivity resulting from massless Dirac fermions. 

Previous experiments on graphite and graphene reported that hot carriers transfer most of their energy to hot optical phonons within 500 fs \cite{PhysRevLett.102.086809,apl/96/8/10.1063/1.3291615,apl/92/4/10.1063/1.2837539,PhysRevLett.95.187403,nl8019399}. They observed a quick buildup of G-phonon occupation through e-ph coupling \cite{srep00064,PhysRevB.80.121403}, but they were not able to clearly resolve the short-time delay region and identify which transition is dominant. Due to finite time resolution, the G-phonons generated directly from e-h recombination are supposed to appear at negative times. Given our 90 fs time resolution, the increase of phonon population would be detected at $\sim$ -35 fs (see the green dashed line in Fig. \ref{Ph_T}). The abrupt appearance of G-phonons near $t$ = 0 therefore suggests that G-phonon emission occurs with a time delay. The presence of a time delay is also clear when the probe pulse is used without the pump. In this  case photons in the early-time part of a probe pulse pump the system, and the photons coming later but still belonging to the same pulse serve as an excitation for Raman scattering (This process is called self-pumping \cite{PhysRevB.80.121403}). Due to the short time duration of the probe pulse ($\sim$50 fs), self-pumping integrates the excitations within the first 35 fs, which suggests that G-phonons are emitted more than 35 fs after a pump pulse, not about 10 fs as expected from direct decay of e-h pairs  into this phonon. 

Another experimental result contradicting the the scenario of direct e-h recombination is the slower G-phonon buildup with lower pump fluence. The opposite is expected, because e-ph coupling was previously reported to increase with decreasing electronic temperature  \cite{PhysRevB.92.184303,PhysRevLett.99.176802}.

Slow thermalization of carriers cannot explain these results. If hot carriers couple to G-phonons directly, G-phonons can be emitted regardless of whether carriers are thermalized or not as long as hot carriers occupy the states near $E_F$. The inverse Auger process that dominates e-e scattering leads to a rapid accumulation of hot carriers near $E_F$ within 30 fs \cite{PhysRevLett.115.086803}.  

The presence of the time delay and the unusual fluence dependence lead us to propose an alternative relaxation pathway where e-h pairs decay primarily into K-point optical phonons. These in turn decay into the G-phonon via strong ph-ph coupling (Fig. \ref{Ph_T}\color{blue}(c)\color{black}). Consistent with this scenario, lifetimes calculated from e-ph coupling for K-phonons are 176 fs, 3 times faster than G-phonons \cite{PhysRevLett.93.185503,PhysRevB.73.155426}. 
Furthermore, strong 4-phonon scattering at high temperature has been reported \cite{PhysRevLett.104.227401,Montagnac201368} with the inferred lifetimes for the decay of K-phonons into G-phonons of about 132 fs (latter based on the G-phonon linewidth of over 40 cm$^{-1}$ at $T$ $>$ 2500 K \cite{Montagnac201368}. The inverse process where K phonons decay to G phonons has the same time constant.). Thus the overall timescale for this indirect process, 300 fs, is consistent with the observed phonon buildup ($\sim$220 fs). The fact that 2D peak is more intense than 2D$'$ peak (double resonance for $\Gamma$ phonons) also suggests a stronger e-ph coupling for K-phonons \cite{nnano.2013.46,PhysRevB.77.041409,PhysRevB.78.081406}.

In order to understand how K-phonons may be responsible for the late appearance of G-phonons, we simulate the optical phonon temperature based on the above scenario where strong anharmonic hot phonons are involved. Our model assumes that carriers thermalize rapidly into a Fermi-Dirac distribution with a well-defined temperature and that G phonons and K phonons thermalize among themselves. The electronic system, G-phonons, and K-phonons are characterized by their temperatures $T_{el}$, $T_{G}$, and $T_{K}$ and are linked by e-ph coupling, $\Gamma_{e-ph}(T_{el},T_{G/K})$, and ph-ph coupling between G and K-phonons,  $\Gamma_{ph-ph}(T_G,T_K)$. This leads to three coupled differential equations:
\begin{align} \label{eq_sim}
\frac{dT_{el}}{dt}&=\frac{I(t)-\Gamma_{e-ph}(T_{el},T_G)-\Gamma_{e-ph}(T_{el},T_K)}{C_{el}(T_{el})},\nonumber\\
\frac{dT_{K}}{dt}&=\frac{\Gamma_{e-ph}(T_{el},T_{K})-\Gamma_{ph-ph}(T_K,T_G)}{C_{K}(T_{K})}-\frac{T_{K}-T_0}{\tau},\nonumber\\
\frac{dT_{G}}{dt}&=\frac{\Gamma_{e-ph}(T_{el},T_{G})+\Gamma_{ph-ph}(T_K,T_G)}{C_{G}(T_{G})}-\frac{T_{G}-T_0}{\tau},
\end{align}
where $I(t)$ is the laser irradiance, $C_{el}$ and $C_{G/K}$ are the heat capacity of electrons and G$/$K-phonons respectively, $T_0$ is room temperature, and $\tau=2.4$ ps  is the measured decay rate of optical phonons into low-energy phonons through anharmonicity. $I(t)$, $\Gamma_{e-ph}(T_{el},T_{G/K})$, and $\Gamma_{ph-ph}(T_G,T_K)$ contain adjustable parameters to fit experimental results (see supplementary for details). Combining ARPES and Raman results constrains the fits so that the same adjustable parameters have to describe the ARPES and the Raman data. Specifically, having to obtain the ARPES simulation close to the experiment constrains $C_{el}$.

The simulation without coupling between G and K phonons ($\Gamma_{ph-ph}$=0) is equivalent to the 2T model. It always gives a curve that is shifted from the Raman data to earlier times (Fig. \ref{Ph_T}\color{blue}(a)\color{black}) because the 2T model predicts that the G-phonons are generated at $t$ $\sim$ -35 fs due to the finite width of the pulse. Note that the coupling constant $\lambda_G$ in $\Gamma_{e-ph}$ only determines the time constant of the G-phonon generation, but it has no influence on when $T_G$ starts to rise. As a result, the phonon dynamics in the early stage of relaxation is clearly not consistent with the 2T model.

We call the simulation with nonzero $\Gamma_{ph-ph}$ the anharmonic hot phonon model (AHP model), because it includes anharmonic couling between the G and K phonons. Due to energy and momentum conservation, the two K-phonons can decay into one G-phonon and one $\Gamma$ phonon in the branch of lower energy, or decay into one G-phonon and one K-phonon in the branch of lower energy (Fig. \ref{dispersion}).  It  agrees much better (Fig. \ref{Ph_T}\color{blue}(a,b)\color{black}) with the data. Within this picture, pump of lower power (0.15 $mJ/cm^2$) gives a slower G-phonon buildup (see Fig. \ref{fluence_dep}) consistent with decreased anharmonicity at lower phonon temperature. So the inclusion of ph-ph scattering is essential for explaining the observed time delay.

Some previous experiments and calculations in the literature agree with out result.  They are not consistent with the dominance of e-h channel in the G-phonon population buildup. For example, phonons are still emitted efficiently in the early stage even when doping puts $E_F$ well below or above half of G-phonon energy \cite{nl301997r}. As the interband transition that contributes to G-phonon emission is suppressed with doping, the G-phonons would not be generated as fast as in the undoped samples by direct e-h recombinations, which suggests that G-phonons are emitted by other processes. However, all these results are consistent with our scenario.

To conclude, we applied TRR and trARPES to investigate e-ph relaxation in graphite within a few hundred femtoseconds. Our simulations combined with experimental results show that instead of electrons coupling to G-phonons directly, K-phonons are emitted via strong e-ph scattering and then transfer energy to G-phonons via ph-ph scattering. This relaxation pathway is also consistent with previously published experiments and calculations \cite{nl301997r,Montagnac201368,PhysRevLett.93.185503,PhysRevB.78.081406}, though the mechanism was not explicitly discussed. 

Our findings have a broader implication for understanding of e-h relaxation processes in other materials such as light absorbers where maximizing photoexcited carrier lifetime is an important engineering challenge. For example in graphite and graphene, carrier recombination could be slowed down significantly by reducing the coupling between the K-point phonons and G-phonons. Thus it is essential to understand and control ph-ph interactions in addition to e-ph ones in order to design better optical absorbers and optoelectronic devices.

\section*{Methods}

Our TRR setup (Fig. \ref{long_time}\color{blue}(a)\color{black}) uses a 20 kHz 790 nm (1.57 eV) laser pulse from an amplified mode-locked Ti:sapphire laser as the pump pulse. The pulse width is about 50 fs. Second harmonic generation at 395 nm (3.14 eV) was used as the probe light source for Raman scattering. The cross-correlation measurement of the pump and probe pulses gave the time resolution of 90 fs. The scattered light was collected by a pair of parabolic mirrors and then analyzed on a McPherson custom triple spectrometer equipped with a water-cooled CCD detector. The spectral resolution is limited by the time-energy Heisenberg uncertainty principle for the probe pulse. The experiment was performed in air at room temperature. The fluence of the pump was from 0.15 $mJ/cm^2$ to 5  $mJ/cm^2$, and the probe fluence was 500 $\mu J/cm^2$. TrARPES used the same laser system with 1.57 eV for the pump with a fluence of 130 $\mu J/cm^2$ and 6.3 eV for the probe (Fig. \ref{long_time}\color{blue}(a)\color{black}). The time resolution for trARPES based on a cross-correlation measurement was 110 fs. 

Time zero in TRR was determined by measuring the sum frequency generation (SFG) intensity of pump and probe pulses on graphite. SFG that originates from the weak nonlinear dielectric polarization of graphite is greatly enhanced when pump and probe pulses arrive at graphite at the same time, which allows us to determine time zero accurately. The SFG intensity at various delay time fits well to a Gaussian function that gives an accuracy of $\pm$4 fs in the determination of time zero. The 0.1 $\mu$m translation resolution of the delay stage only adds 0.67 fs to the time uncertainty. Therefore the total uncertainty of time zero is less than 5 fs.

\bibliography{report_reference}

\begin{thebibliography}{18}
\bibitem [{foo()}]{footnote}%
  \BibitemOpen
  \href@noop {} {\ }\bibinfo {note} {In principle the K phonon population
  should also affect the 2D peak intensity. However under the excitation energy
  of 3 eV, the momentum of phonons involved in the double resonance process is
  actually further from K points than those phonons emitted during the electron
  relaxation, i.e. they are different phonon modes. Since photoexcied electrons
  thermalize quickly \cite{nphys2564}, most phonons are emitted from the hot
  electrons near the Fermi level and therefore have a momentum near K or
  $\Gamma$ points. The absence of the 2D peak in the anti-Stokes side also
  supports that no phonons are generated away from K points, suggesting that
  the effect of the hot phonon population on the 2D peak is
  negligible.}\BibitemShut {Stop}%
\end{thebibliography}

\section*{Acknowledgements}

The time-resolved Raman work at the University of Colorado was supported by the NSF under Grant No. DMR-1410111. The time-resolved ARPES work was supported by NSF grant DMR-1508785, with the equipment purchased in part through NSF MRI-1546961. We thank Justin Griffiths, Justin Waugh, Tom Nummy, Haoxiang Li, and Kyle Gordon for help with the experiments.
\section*{Author contributions statement}

D.R. conceived the project. J.Y. performed the Raman experiments and analyzed the Raman data. S.P. performed the ARPES experiments and analyzed the ARPES data. J.Y. wrote the paper with D.R. J.Y. performed the temperature simulation. D.D. and D.R. supervised the project. All authors contributed to the discussions and manuscript preparation.

\section*{Additional information}
\textbf{Competing financial interests:} The authors declare no competing financial interests.


\end{document}


\title{Supplementary: Novel Electron-Phonon Relaxation Pathway in Graphite Revealed by Time-Resolved Raman Scattering and Angle-Resolved Photoemission Spectroscopy}
\author{Jhih-An Yang}
\affiliation{Department of Physics, University of Colorado at Boulder, Boulder, CO 80309}
\author{Stephen Parham}
\affiliation{Department of Physics, University of Colorado at Boulder, Boulder, CO 80309}
\author{Daniel Dessau}
\affiliation{Department of Physics, University of Colorado at Boulder, Boulder, CO 80309}
\affiliation{Center For Experiments on Quantum Materials, University of Colorado, Boulder, CO 80309}
\author{Dmitry Reznik}
\affiliation{Department of Physics, University of Colorado at Boulder, Boulder, CO 80309}

\maketitle

\section{Details of simulation}

Here we discuss the details of the simulation of optical phonon temperature by Eq. 1 in the main text. We used the form $I(t)=(F/2\tau_p)sech^2(t/\tau_p)$ to model the pulse, where $F$ is the absorbed fluence and $\tau_p$ is the time duration. We use $\tau_p=33.15$ fs (FWHM=58.44 fs) so the crosscorrelation width is 90 fs.

The electronic heat capacity can be expressed as:
\begin{equation}
C_{el}(T_{el})=\frac{\mathrm{d}}{\mathrm{d}T_{el}}\int D(\epsilon)f(\epsilon,T_{el})\epsilon\mathrm{d}\epsilon,
\end{equation}
where $D(\epsilon)$ is the density of states for the linear dispersion of bands, and $f(\epsilon,T_{el})=[\mathrm{exp}(\epsilon/kT_{el})+1]^{-1}$ is the Fermi-Dirac distribution by assuming zero chemical potential for simplicity.

The heat capacity for G and K phonons can be expressed as:
\begin{equation}
C_{G/K}=\frac{\mathrm{d}}{\mathrm{d}T}\frac{\hbar\Omega}{\mathrm{exp}(\hbar\Omega/kT)-1}N,
\end{equation}
where $\hbar\Omega$ is the phonon energy and $N$ is the number of phonon modes near $\Gamma$ or K points. For further simplicity we consider G-phonons and K-phonons having the same energy of 200 meV (in reality $\hbar\Omega_G=200$ meV and $\hbar\Omega_K=160$ meV).

In our model we used anharmonic decay rate $\tau=2.4$ ps based on the lifetime of G-phonons measured by our experiment. We also assumed the same decay rate for K-phonons.

We constructed the e-ph decay rate by considering possible transitions. The G-phonons can be emitted only by interband intravalley scattering of electrons due to energy and momentum conservation. Both intraband intervalley and interband intervalley transitions can contribute to the emission of phonons near K points. Note that the phonons from intraband intervalley transitions are not exactly at K points \cite{PhysRevB.84.035433}. In general we should integrate all possible states that contribute to the K-phonon emission to obtain the energy transfer rate. Since the main purpose of this model is to explain how K-phonons are responsible for the time delay of G-phonons, the detail of e-ph scattering for K phonons is beyond this scope. In fact how electrons couple to K phonons has a minor influence on the time delay as long as the coupling is strong. Therefore we only considered interband intervalley transitions for K-phonons in our model for simplicity. The exchange rate $\Gamma_{e-ph}$ for G/K-phonons consequently can be expressed as \cite{PhysRevLett.59.1460}:
\begin{align}
&\Gamma_{e-ph}=\lambda_{G/K}\left\{f(\frac{\hbar\Omega}{2},T_{el})\left[1-f(\frac{-\hbar\Omega}{2},T_{el})\right](1+n_{G/K})\right.\nonumber\\
&\left.-f(\frac{-\hbar\Omega}{2},T_{el})\left[1-f(\frac{\hbar\Omega}{2},T_{el})\right]n_{G/K}\right\}D(\frac{\hbar\Omega}{2})D(\frac{-\hbar\Omega}{2}),
\end{align}
where $n_{G/K}=[\mathrm{exp}(\hbar\Omega/kT_{G/K})-1]^{-1}$ is the G$/$K-phonon population and $\lambda_{G/K}$ is the e-ph coupling constant. 

The energy exchange rate between G and K phonons due to 4-phoonon scattering can be expressed as:
\begin{align}
&\Gamma_{ph-ph}(T_K,T_G)=b[n_K^2(1+n_G)^2-(1+n_K)^2n_G^2],
\end{align}
where $b$ is ph-ph coupling strength.

To solve the coupled differential equations, we used the Euler method \cite{butcher2008numerical}. Starting with the initial condition $T_{el,0}$, $T_{G,0}$, and $T_{K,0}$, we take a small time step $\Delta t$ and evaluate the temperatures $T_{el,1}$, $T_{G,1}$, and $T_{K,1}$ at next time $t_1=t_0+\Delta t$. That is,
\begin{align}
T_{el,n+1}&=T_{el,n}+\nonumber\\
&\Delta t \frac{I(t_n)-\Gamma_{e-ph}(T_{el,n},T_{G,n})-\Gamma_{e-ph}(T_{el,n},T_{K,n})}{C_{el}(T_{el,n})},\nonumber\\
T_{G,n+1}&=T_{G,n}+\nonumber\\
&\Delta t\left( \frac{\Gamma_{e-ph}(T_{el,n}T_{G,n})+\Gamma_{ph-ph}(T_{K,n},T_{G,n})}{C_{G}(T_{G,n})}-\frac{T_{G,n}-T_0}{\tau}\right),\nonumber\\
T_{K,n+1}&=T_{K,n}+\nonumber\\
&\Delta t \left(\frac{\Gamma_{e-ph}(T_{el,n},T_{K,n})-\Gamma_{ph-ph}(T_{K,n},T_{G,n})}{C_{K}(T_{K,n})}-\frac{T_{K,n}-T_0}{\tau}\right)
\end{align}
From the temperatures at $t_{n}$, we can evaluate the temperatures at $t_{n+1}$ and so on. The simulated results are convolved with the instrument response that accounts for the finite width of the probe pulse.

The initial condition was set to be room temperature at negative time. The number of phonon modes $N$ in the initial condition is determined by the experimental electronic and phonon temperatures at small positive times. The larger the difference between $T_{el}$ and $T_G$ is, the larger $C_{G/K}$ (and larger $N$) is. This is because smaller $C_{G/K}$ leads to a larger change in temperature and higher equilibrium temperature the whole system reaches.

\begin{figure}[t]
\includegraphics[width=0.5\textwidth]{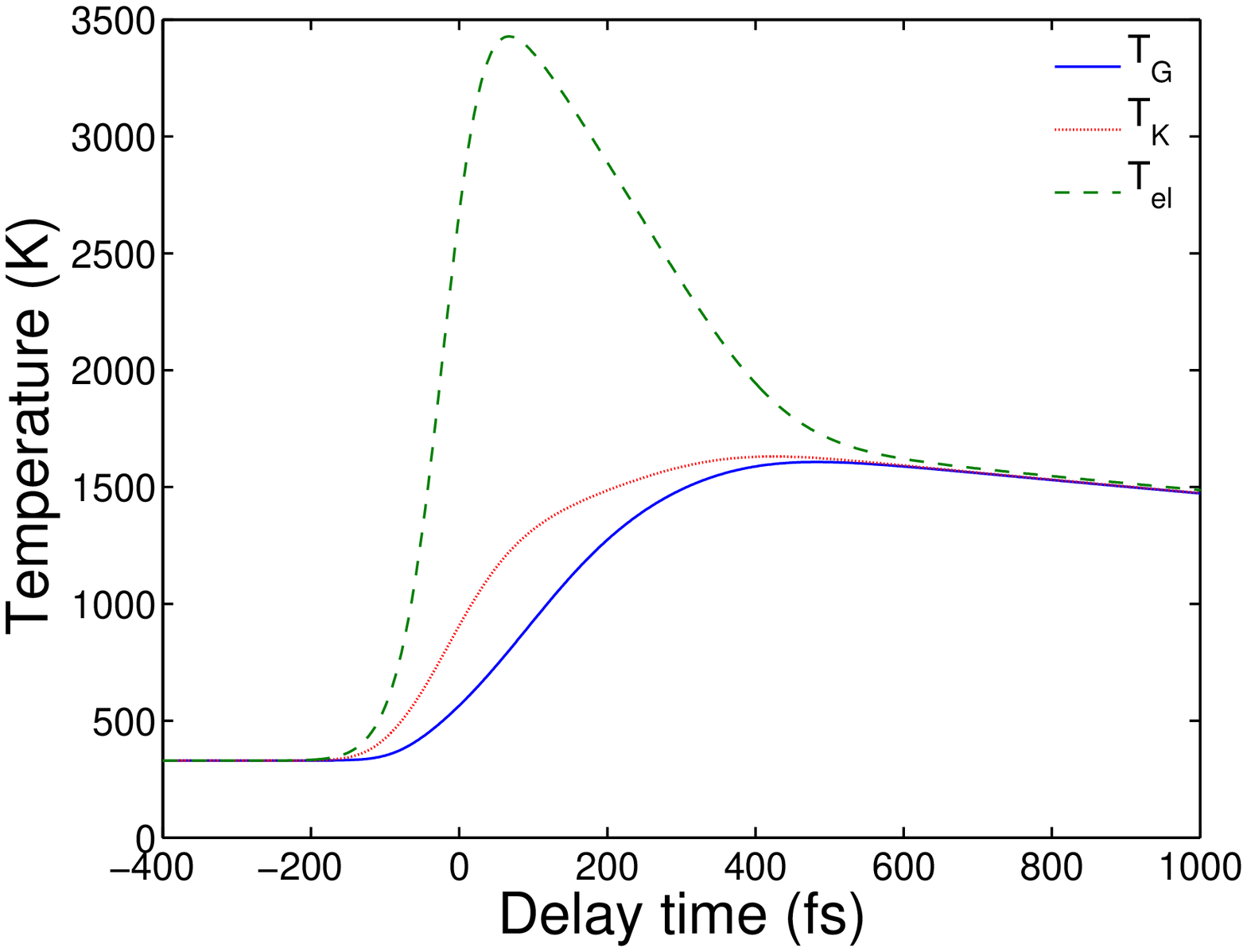} 
\caption{G-phonon temperatrue, K-phonon temperature, and electronic temperature as a function of time from the simulation. \label{3Tmodel}}
\end{figure}

$F$, $\lambda_G$, $\lambda_K$, and $b$ are adjustable parameters in the simulation to fit experimental results. Here we focus on the region at $t\leq800$ fs because this is the region where we took most of the Raman data. The long decay of the phonon population has been characterized elsewhere \cite{PhysRevB.80.121403}. It can be well described by a single exponential (see the inset of Fig. 2(e) in the main text) due to anharmonic coupling to acoustic phonons. The solution of Eq. 1 in the main text actually follows a single exponential function at long delay time as well. The simulated results are shown in Fig. \ref{3Tmodel}. The parameters for the best fit are $F$ =  0.0029 $mJ/cm^2$, $\lambda_G=2.29\times10^{-5}$ $eV^2m^2/s$, $\lambda_K=2\times10^{-4}$ $eV^3m^2/s$, and $b=2.01\times10^{30}$ $eV/m^2s$. The parameter $F$ is about 2\% of the measured pump fluence 0.15 $mJ/cm^2$, consistent with 2.3 \% for single layer absorption. The $\chi^2$ ($\chi^2=\sum(T_{exp}-T_{model})^2/\sigma_i^2$, where $T_{exp}$ is the experimental temperature, $T_{model}$ is the simulated temperature, and $\sigma$ is the errorbar) for the 2T model ($b=0$) is 76, while the $\chi^2$ for the AHP model ($b\neq0$) is 33.5. The difference of the $\chi^2$ mainly comes from the deviations from $t=-40$ fs to $t=300$ fs.

Our model qualitatively describes the role of K-phonons and how they are responsible for the late appearance of G-phonons. We did not consider e-ph scattering for other phonons as their coupling strengths are much less than of G or K phonons.

\section{2D peak intensity and its relation to electronic temperature}
\begin{figure}[t!]
\includegraphics[width=0.5\textwidth]{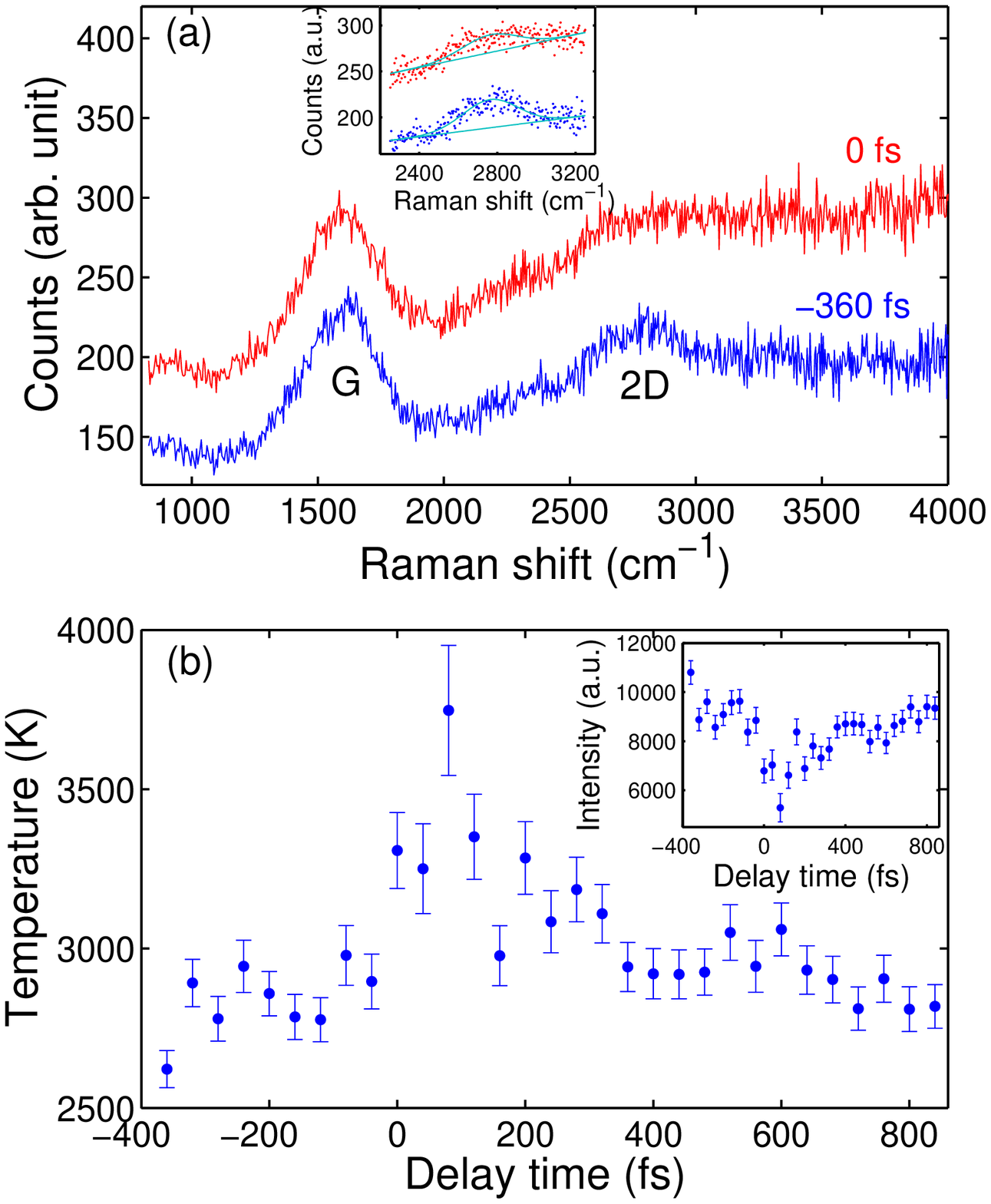} 
\caption{(a) Stokes Raman spectra at t = -360 fs and t = 0 fs. Inset: Region of the 2D peak with fits to the data (see text). (b) Electronic temperature extracted from Eq. (\ref{2DT}). The high electronic tempeature at negative times is due to self-pumping by the probe. Inset: Integrated intensity of the 2D peak on the Stokes side as a function of delay time.  \label{2D_temperature} }
\end{figure}

The 2D peak intensity on the Stokes side directly measures electronic temperature because it decreases with increasing temperature \cite{PhysRevB.58.5435,apl1.123096,jp809501e} or doping \cite{nnano.2008.67,PhysRevB.79.155417} . More electrons in the conduction band enhance the e-e scattering rate and tend to suppress the double resonance responsible for the 2D peak \cite{PhysRevB.84.035433,PhysRevB.80.165413,PhysRevB.78.125418}. In our experiment the 2D peak intensity drops quickly around time zero and then recovers more gradually (Fig. \ref{2D_temperature}). 
Here we used the relation of 2D peak intensity and the electronic linewidth in Ref. \cite{PhysRevB.84.035433} since e-e scattering dominates at high temperature \cite{nphys2564,footnote}. The e-e scattering rate is proportional to temperature \cite{PhysRevB.83.155441}, so we can rewrite the relation as:
\begin{equation} \label{2DT}
I_{2D}=\frac{a'}{\gamma^{2}}=\frac{a}{T_{el}^{2}} \text{,}
\end{equation}
where $a$ and $a'$ are constants and $\gamma$ is the electronic linewidth. We first fit the 2D peak with a Gaussian, whose width was fixed at the energy resolution, plus a linear background (Fig. \ref{2D_temperature}\color{blue}(a) \color{black} inset). The background should be fit with Planck's law \cite{PhysRevLett.105.127404}, but in a small energy region a straight line is a good fit. Since electrons and G-phonons have the same temperature at t $>$ 500 fs, we found the proportionality $a$ to be $7.42\times10^{10}$ from the intensities at t $>$ 500 fs. We calculated the electronic temperature based on the changes in intensity for t $<$ 500 fs (Fig. \ref{2D_temperature}\color{blue}(b)\color{black}). $T_{el}$ extracted from the 2D peak increases rapidly after an optical excitation and reaches 3800 K at t = 80 fs. Compared to trARPES results, the measurement of the 2D peak differs from the real $T_{el}$. This is because the e-e scattering rate might not follow the simple relation Eq. \ref{2DT}. Self-pumping also causes a deviation from real electronic temperature. As can be seen in Fig. \ref{3Tmodel}, $T_e$ is already over 2000 K at $t=0$ fs. This explains why $T_e$ at negative times measured from the 2D Raman peak is always above 2500 K, since self-pumping integrates the excitations within about the first 35 fs. This prevents us from extracting real $T_e$ using the 2D peak. Nevertheless, the 2D peak still provides qualitative information about $T_e$. There is no such a problem for measuring the G phonon population since we did not observe any self-pumping in G phonons.

\section{Laser profile}
\begin{figure}[t]
\includegraphics[width=0.5\textwidth]{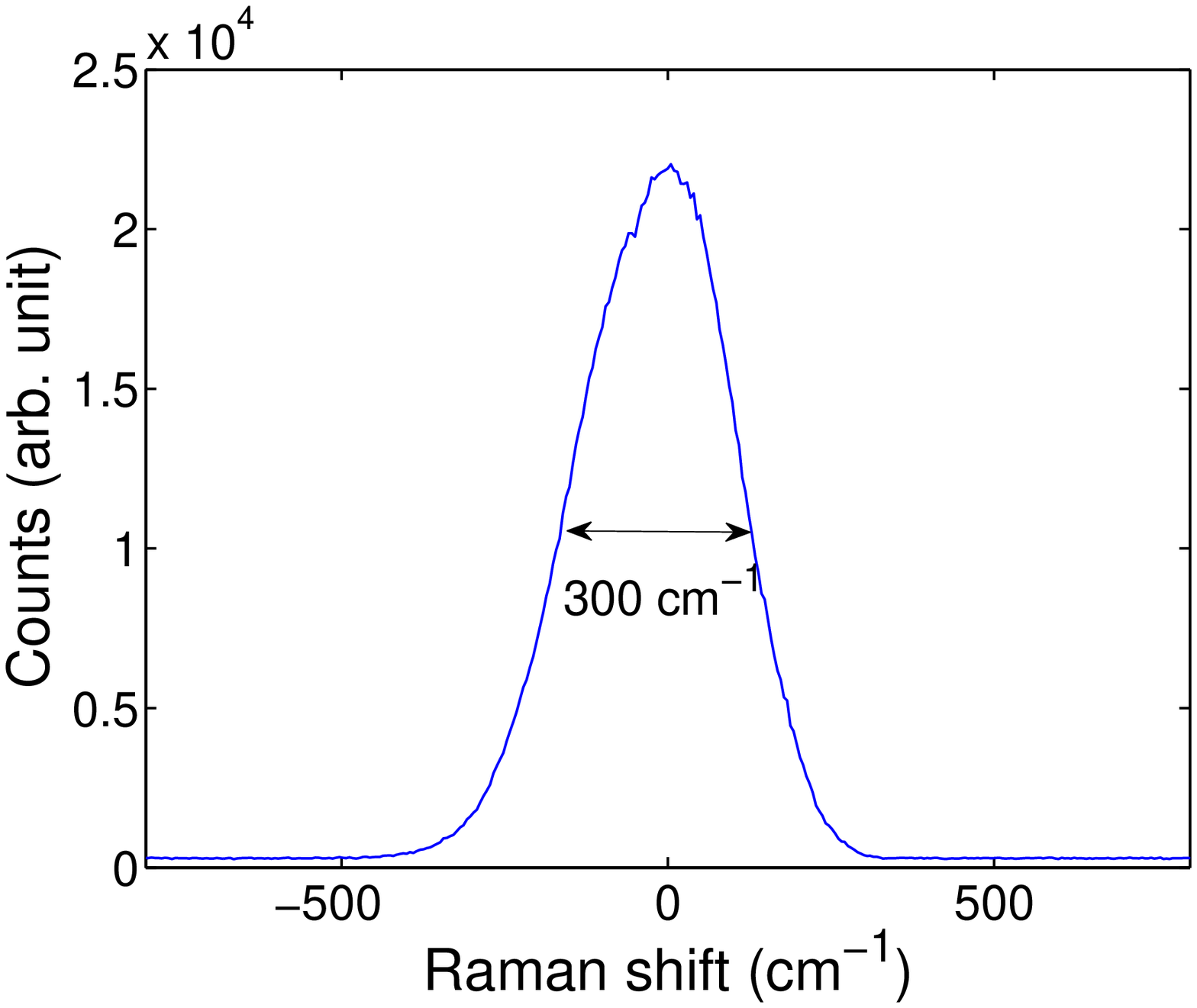} 
\caption{The energy profile of a probe pulse. The width is about 300 $cm^{-1}$.\label{laser}}
\end{figure}

The frequency profile of a probe pulse is shown in Fig. \ref{laser}. The full width at half maximum (FWHM) is about 300 $cm^{-1}$. Now we exam whether the temporal profile is the Fourier-transform-limit. Suppose the laser is described by a Gaussian function in the frequency domain 
\begin{equation}
f(E)=e^{-\frac{E^2}{2c^2}}
\end{equation}
with FWHM$=2\sqrt{2log2}c$. Its Fourier transform is 
\begin{equation}
F(t)=\frac{c}{\sqrt{2\pi}\hbar}e^{-\frac{c^2t^2}{2\hbar^2}}.
\end{equation}
Therefore the FWHM in the time domain is $2\sqrt{2log2}\hbar/c$. The product of the widths (FWHM) in the time and frequency domains is:
\begin{equation}
\Delta t\Delta E=\left(2\sqrt{2log2}c\right)\left(\frac{2\sqrt{2log2}\hbar}{c}\right)=8log2\hbar
\end{equation}
Here $\Delta E=300$ $cm^{-1}$, so we get $\Delta t=$ 49 fs. This confirms that our pulse is close to the Fourier limit, given that the measured crosscorrelation width between pump and probe pulses is 90 fs.

\section{Nonthermal carriers at early times}
\begin{figure}[t!]
\includegraphics[width=0.5\textwidth]{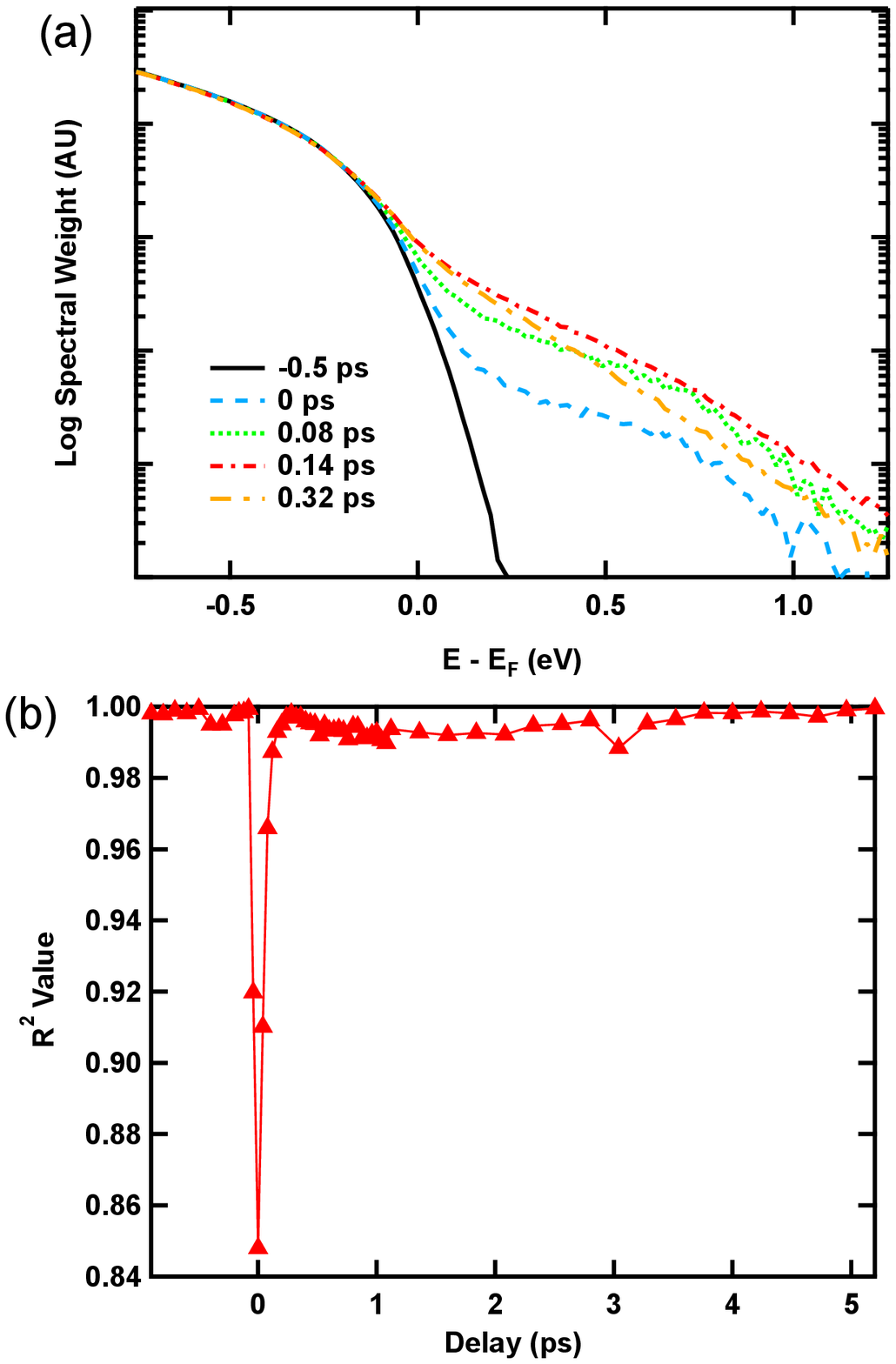} 
\caption{(a) Momentum-integrated ARPES intensity at early delay times. (b) $R^2$ value of the fittings at various delay times. (see text) \label{nonthermal}}
\end{figure}
Non-thermal behavior of photo-excited carriers is observed in graphite during initial excitation. Fig. \ref{nonthermal}\color{blue}(a) \color{black} shows the momentum-integrated ARPES intensity, which is indicative of the occupation function (above $E_F$). Here we see qualitatively that the $t$ = 0 and 0.08 ps traces (gray) do not follow a Fermi-Dirac distribution, which should appear linear above $E_F$ on a log-linear plot. This is shown quantitatively in Fig. \ref{nonthermal}\color{blue}(b) \color{black} as a reduction of the $R^2$ value of the fitting for time delays near t = 0 (fits are shown in the main text). For $t>$ 0.12 ps the system has effectively thermalized, consistent with previous reports \cite{PhysRevLett.115.086803,nmat3757} and our experimental time resolution of 0.11 ps.

\bibliographystyle{apsrev4-1}
\bibliography{report_reference}